\documentclass[aps,pre,twocolumn,showpacs,preprintnumbers,longbibliography,superscriptaddress]{revtex4-1}
\usepackage{amsmath}
\usepackage{times,mathptmx,amssymb}
\usepackage{graphicx}
\usepackage{xcolor}
\usepackage{siunitx}

\newcommand{\derpar}[2]{\frac{\partial #1}{\partial #2}}
\newcommand{\VEC}[1]{\mathbf{#1}}

\begin{document}
\title{Modelling of active contraction pulses in epithelial cells using the vertex model}

\author{Fernanda P\'erez-Verdugo}
\affiliation{Departamento de F\'{\i}sica, FCFM, Universidad de Chile, Santiago, Chile}

\author{Germ\'an Reig}
\affiliation{Escuela de Tecnolog\'\i a M\'edica y del Centro Integrativo de Biolog\'\i a y Qu\'\i mica Aplicada (CIBQA), Universidad Bernardo O'Higgins, Santiago, Chile}

\author{Mauricio Cerda}
\affiliation{Institute of Biomedical Sciences, Faculty of Medicine, Universidad de Chile, PO Box 70031, Santiago, Chile}
\affiliation{Biomedical Neuroscience Institute, Independencia 1027, Santiago, Chile}
\affiliation{Center for Medical Informatics and Telemedicine (CIMT), Facultad de Medicina, Universidad de Chile, Santiago, Chile}

\author{Miguel L. Concha}
\affiliation{Institute of Biomedical Sciences, Faculty of Medicine, Universidad de Chile, PO Box 70031, Santiago, Chile}
\affiliation{Biomedical Neuroscience Institute, Independencia 1027, Santiago, Chile}
\affiliation{Center for Geroscience, Brain Health and Metabolism, Santiago, Chile}

\author{Rodrigo Soto}
\affiliation{Departamento de F\'{\i}sica, FCFM, Universidad de Chile, Santiago, Chile}

\begin{abstract}
Several models have been proposed to describe the dynamics of epithelial tissues undergoing morphogenetic changes driven by apical constriction pulses, which differ in where the constriction is applied, either at the perimeter or medial regions. To help discriminate between these models, using the vertex model for epithelial dynamics, we analysed the impact of where the constriction is applied on the final geometry of the active cell that is reducing its apical size. We find that medial activity, characterised by a reduction in the reference area in the vertex model, induces symmetry breaking and generates anisotropic cell shapes, while isotropic cell shapes and larger contractions occur when the reference perimeter in the model is reduced. When plasticity is included, sufficiently slow processes of medial contractile activity, compared with typical apical constriction pulses, can also achieve significant cell contraction. Finally, we apply the model to describe the active apical contractile pulses observed during cellular mitotic events within the epithelial enveloping cell layer in the developing annual killifish Austrolebias nigripinnis, being able to quantitatively describe the temporal evolution of cell shape changes when perimeter activity and area plasticity are included. A global fit of all parameters of the vertex model is provided. 
\end{abstract}

\maketitle

\section{Introduction.}
Our view about epithelia has changed dramatically over the years from static and rigid cellular structures functioning as simple mechanical barriers to highly dynamic supra-cellular arrangements of polarised cells actively involved in morphogenetic processes (reviewed in \cite{solnica2005conserved,quintin2008epithelial}). Examples of morphogenesis involving dynamic epithelia extend to many living organisms and include, among others, germ band elongation and gastrulation in \textit{Drosophila} \cite{martin2020physical,he2010tissue}, neural tube formation in \textit{Xenopus laevis} \cite{christodoulou2015cell}, and convergent and extension movements in teleost fish (reviewed in \cite{solnica2005conserved}). Cells within epithelia are maintained in close contact with each other by cell-cell adhesion complexes, and the dynamic regulation of the cytoskeleton mold their shape and behaviour in response to both external and internal biomechanical factors (reviewed in \cite{gumbiner1996cell,takeichi2014dynamic}). Among the most relevant epithelial cell shape changes that promote tissue remodelling in a wide range of homeostatic and developmental contexts is apical constriction, the process by which the apical surface of the cell contracts, causing the cell to take on a wedged shape (reviewed in \cite{martin2014apical}). At the molecular level, apical constriction is regulated primarily by the contraction and flows of actomyosin networks present at the apical side of the cell. Though this mechanism is operating at a single cell level, it has been documented that events of apical constriction are highly coordinated at the supra-cellular level within epithelia. In accordance, forces generated by apical constriction are transmitted to surrounding cells through cell-cell adhesion complexes and contribute to significant macroscopic deformation of tissues  \cite{gorfinkiel2016actomyosin}. Conversely, the mechanical environment imposed by neighbouring cells can regulate the dynamics of apical constriction \cite{martin2014apical}. 

Apical constriction can be continuous or pulsed (reviewed in Refs.~\cite{miao2020pulse,sutherland2020pulsed}) and different force-dependent mechanisms have been proposed to drive this process, including the purse-string, meshwork, and ratchet models \cite{martin2010pulsation}. In the purse-spring model, the contraction force is generated by an actomyosin ring localised in the internal perimeter of a cell, such as the resultant forces are aligned with the sides of the cell. In the meshwork model, the medial cortex of actomyosin suffers a constriction in the first place. As this actin meshwork is connected to adherens junction sites in the cellular borders, the result is an inward force that pulls these sites into the cell, constricting it. Finally, in the ratchet model, the net cellular constriction is achieved from a sequence of discrete steps of contraction of the medioapical actomyosin network followed by a stabilisation phase, leading to pulse disassembly. Because of the peripheral and medial localisation of the purse string and meshwork mechanisms, it has been proposed that they might be most effective at generating isotropic and anisotropic tensile forces, respectively  \cite{martin2010pulsation}.  However, there is no complete characterisation of the possible scenarios for the evolution of tissues when the different models are used to describe the apical constriction  events.

To describe and compare the apical constriction events produced either by perimeter or medial activity, we characterise the geometry of the active cell after the pulse, which is an observable that is easily accessible in experiments and, therefore, can be used to discriminate between the models. To study the cell geometries, we 
 use the vertex model, which has been extensively applied to describe the dynamics of epithelial tissues \cite{nagai1988vertex, nagai2001dynamic, inoue2016mechanical}. In this mathematical model, a thin epithelium, in which the cell heights do not change appreciably, is considered as a two-dimensional tessellation of polygonal cells. The degrees of freedom of the model are the cell vertices, which evolve to minimise a free energy functional. In accordance, we include cell activity in the medial and the perimeter regions of a cell, mimicking the meshwork and the purse-string models, and characterise the final shape of this targeted cell in terms of the net achieved contraction and anisotropy. We do not study the case of contractions achieved by the ratchet model. We find that perimeter activity is the most efficient at favouring intense apical constriction rendering more circular apical cell shapes than medial activity. The latter, in contrast, promotes less intense apical constriction and the apical side adopts a more elongated shape. These findings might indicate that to achieve a significant contraction in systems where the medial region provides the active contractile force, then a ratchet pulsated dynamic is needed. However, if the geometrical evolution is smooth, then the analysis suggests that a perimeter activity is more suitable, in agreement with examined cases in Ref.~\cite{martin2010pulsation}.

\begin{figure}[htb]
\includegraphics[angle=0,width=\columnwidth]{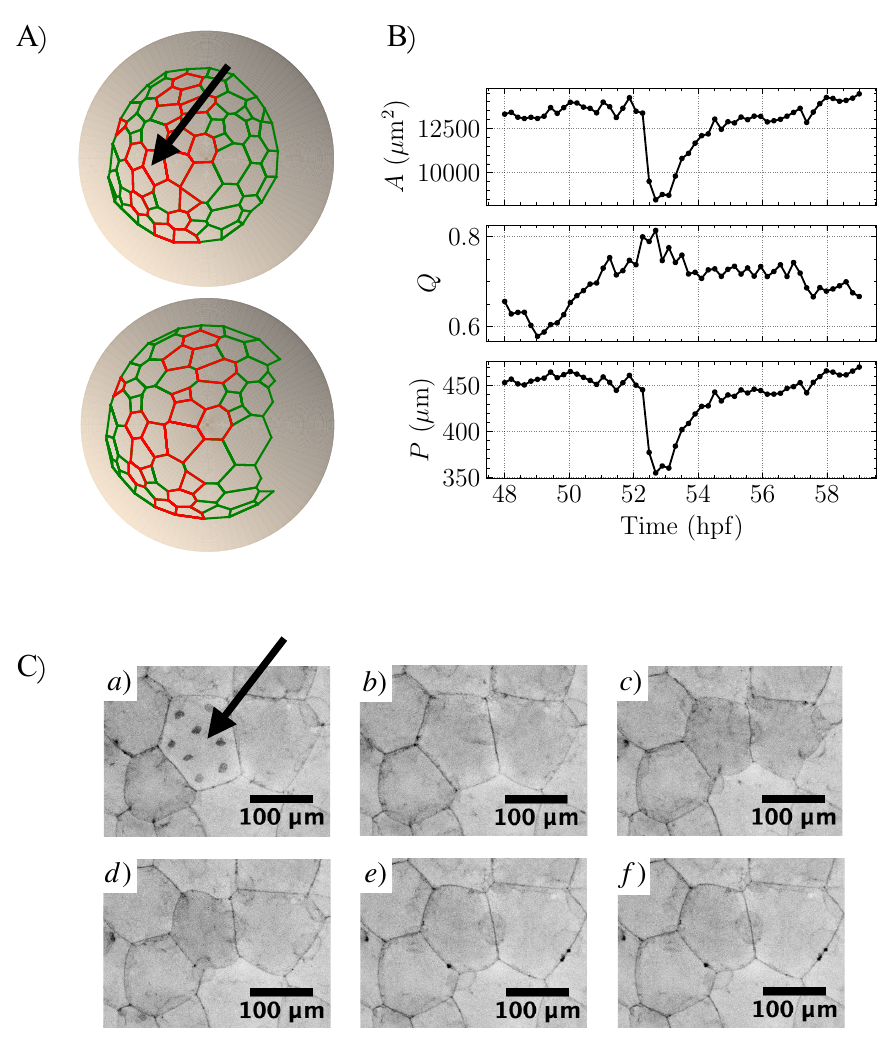}
\caption{A) Temporal progression of epithelial cell shape changes during blastula stages of \textit{Austrolebias nigripinnis} from 48 to \SI{59}{hpf}. Epithelial polygonal EVL cells are denoted in green and those with notable constriction through the timescale considered are shown in red.  B) Temporal evolution of cell shape changes referring to cell area ($A$, top), anisotropy ($Q$, middle) and perimeter ($P$, bottom) that characterise apical constriction in the target cell marked by a black arrow in A) and C). The parameter anisotropy was estimated as $Q = b/a$, where $a > b$ are the principal semi-axis of the ellipse that better fit to the cell (for more details see Ref.~\cite{perezinstability}). C) Representative confocal microscopy-derived planar projections images of single EVL cells suffering transient events of apical constriction. The duration of the complete image sequence of cell shape changes is 3h and has been divided into three major steps including pre-constriction a)-b); active constriction c)-d) and relaxation period leading to shape recovery e)-f). A hallmark of the pre-constriction step is the accumulation of F-actin in nuclei (increase in nuclear fluorescence indicated by an arrow). Images shown in C) were captured using a representative \textit{A.\ nigripinnis} embryo microinjected with lifeact-GFP \cite{riedl2008lifeact}.}
\label{fig.huevo_c22}
\end{figure}

In vivo observations using developing embryos of the annual killifish \textit{Austrolebias nigripinnis} (\textit{A.\ nigripinnis}) have uncovered a series of transient and apparently unsynchronised events of apical constriction within the epithelial enveloping cell layer (EVL) at the blastula stage \cite{reig2017extra}. These constrictions are characterised by an initial phase of fast and short contraction followed by a relaxation period in which the original apical shape and size are recovered. As these apical constriction events are associated with a duplication of the number of nuclei per cell it has been proposed they correspond to events of cytokinetic mitotic failures \cite{reig2017extra}. One of such apical constriction events of EVL cells is shown in Fig.\ \ref{fig.huevo_c22}. Using the EVL of annual killifish at the blastula stage for modelling apical constriction has at least two major advantages in comparison with other developing epithelia. First, at this phase the EVL is composed of a small and fixed number of cells of considerable size. Second, no major morphogenetic movements occur during the events of apical constriction.
These apical constriction events are therefore suitable for the aforementioned analysis, finding that it is indeed possible to discriminate between the models and that the events are described by activity only in the perimeter of the cells.

The article is organised as follows. The vertex model and the method used to consider the cell activity is described in Sect.\ II. Analytic calculations for a single isolated cell are presented in Sect.\ III. Section IV analyses the case of larger tissue in which the vertex model is solved numerically. The comparison of the model with the experiments in \textit{A.\ nigripinnis} is described in Sect.\ V. Finally, in Sect.\ VI, the main conclusions and discussions are presented.

\section{Vertex model}\label{sec.model}
We use a general vertex model in two dimensions, with the energy functional given by
\begin{align}
E = \frac{K_A}{2}\sum_c\left( A_c - A_{0c}\right)^2 + \frac{K_P}{2}\sum_c\left( P_c -
P_{0c}\right)^2+J\sum_{\langle i,j \rangle} l_{ij}.
\label{eq.Evertex}
\end{align}
The first and second sums are over the cells of the system, while the third is over the pair of adjacent vertices $i$ and $j$. $A_c$ and $P_c$  are the area and perimeter of the cell $c$, respectively, and $l_{ij}$ is the length of the cellular edge between the adjacent vertices $i$ and $j$. $A_{0c}$ and $P_{0c}$ are the area and perimeter reference values, to which the cells tend to evolve. Throughout this work, we assume that these quantities are given initially by the values of the area and perimeter of each $c$ cell in $t=0$. The first and second terms of Eq.~\eqref{eq.Evertex} are related to the medial and perimeter actomyosin contractility, respectively; the third one is an adhesion energy term.  

The position of the vertex $\VEC r_i$ follows a variational dynamic, 
\begin{align}
\frac{d \VEC r_i}{dt} = -\gamma \derpar{E}{\VEC r_i},
\label{eq.variationaldyn}
\end{align}
where $\gamma$ is a mobility that we absorb in the elastic parameters of Eq.~\eqref{eq.Evertex}.

The model considers three elastic parameters $K_A, K_P$, and $J$. Here,  we use the dimensionless parameters, $p=K_P/(a^2K_A)$ and $j=J/(a^3 K_A)$, where $a$ is the side of the cells. These parameters quantify the competition between the characteristic time of the apical actomyosin meshwork elasticity and the ones of the perimeter actomyosin and adhesion elasticity, respectively. 
The third term of Eq.~\eqref{eq.Evertex} can be absorbed in the second one by redefining $P_{0c}$ as $P_{0c} - J/(2 K_P)$. Commonly two choices are made, i) setting $P_{0c}=0$ \cite{staple2010mechanics} or ii) setting $J=0$ \cite{nestor2018}. 
However, to do so,  the whole set of $P_{0c}$ must remain constant during the evolution, which is not the case in the present study where cells contract and present plasticity. Hence, we decide to keep all terms, where each one has a clear biophysical interpretation. Also,  $J$ can be thought of as a permanent and homogeneous cortical cell activity.

The cell contractile activity is included by incorporating variations of the reference quantities of a target cell, such that $A_{0c}\rightarrow \left(1-\alpha_A\right)A_{0c}$ or $P_{0c}\rightarrow \left(1-\alpha_P\right)P_{0c}$, while for the rest of the cells $A_{0c}$ and $P_{0c}$ keep their initial value. The dimensionless parameters $\alpha_A$ and $\alpha_P$ represent the activity in the medial and perimeter cellular regions, respectively.

\section{Isolated active hexagonal cell}  \label{sec.17cell}

As the first case of study, we consider a single hexagonal cell of side $a$, initially in its regular shape, with the vertices positions given by $\VEC r^{[0]}$, and for which its equilibrium area and perimeter are given by $A_0=3\sqrt{3}a^2/2$ and $P_0=6a$, respectively.
We consider the vertex model as expressed in Eq.~\eqref{eq.Evertex}, where the elastic coefficients $K_A$ and $K_P$ are assumed to be positive. The parameter $J$ is limited to take positive values to ensure stability  \cite{perezinstability}. We analyse the cases of cell activity with $\alpha_A=0.5$ or $\alpha_P=0.5$ and characterise the response when we allow the cell to deform. These activity values are consistent in the order of magnitude with experiments that measure the changes of actin and myosin during pulsed contractions \cite{martin2009pulsed, mason2016rhoa, michaux2018excitable}.

Then, we perform an affine transformation with respect to the center of the cell by changing the vertex positions as  $\VEC  r^{[0]}_i \rightarrow \left(I+ U\right) \VEC r^{[0]}_i$, where $U$ is a general $2\times2$ matrix of components $u_{ik}$, characterising the deformation.
Expressing $U$  as a linear combination of four deformation modes,
\begin{align}
U_1 &=\begin{pmatrix}
-1 & 0\\
0 & 1
\end{pmatrix} \text{[deviatoric]},  &
U_2 &= \begin{pmatrix}
0 & 1\\
1 & 0
\end{pmatrix} \text{[pure shear]}, \label{eq.modes}\\
U_3 &= \begin{pmatrix}
0 & -1\\
1 & 0
\end{pmatrix} \text{[rotation]},&
U_4 &= \begin{pmatrix}
-1 & 0\\
0 & -1
\end{pmatrix} \text{[contraction]},\nonumber
\end{align}
as $U=\sum_{i=1}^4  v_i U_i$, the variation of the energy can be written as $\Delta E = \hat{E} e(v_1, v_2, v_3, v_4 )$, where $\hat{E} =  {K_A} A_{0}^2/2$ is the energy scale and $e$ is a nonlinear but analytic function, which can be directly obtained replacing the deformation in Eq.\ \eqref{eq.Evertex}.

\begin{figure}[ht!] 
 \includegraphics[width=1\linewidth]{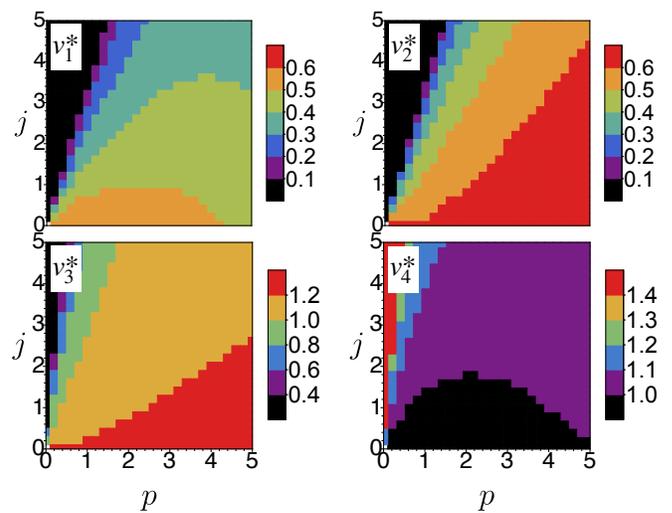} 
\caption{Amplitude of each deformation mode $v_i^*$ that appear as a response of an isolated regular hexagonal cell under the modification of the equilibrium area, with $\alpha_A=0.5$ and  $\alpha_P=0$.}
\label{fig.onecellA0}
\end{figure}

\begin{figure}[ht!] 
 \includegraphics[width=1\linewidth]{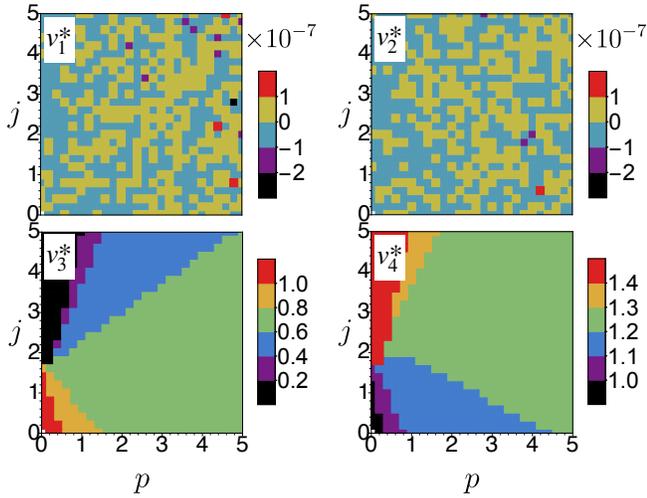} 
\caption{Same as in Fig.\ \ref{fig.onecellA0}, when the equilibrium perimeter is changed, with $\alpha_A=0$ and $\alpha_P=0.5$.}
\label{fig.onecellP0}
\end{figure}

We calculate the set of amplitudes $\left(v_1^*,v_2^*,v_3^*,v_4^*\right)$ that minimise the energy variation. Figures \ref{fig.onecellA0} and ~\ref{fig.onecellP0} show the amplitude of each deformation mode as a response to cell activity, for the cases $\alpha_A=0.5$ (with $\alpha_P=0$) and $\alpha_P=0.5$ (with $\alpha_A=0$), respectively.
The amplitude of the deviatoric and pure shear modes, which are responsible for breaking the isotropy of the system, are null in the case of activity in the perimeter. Hence, cell activity in the inner cellular border might generate isotropic cell shapes, while cell activity in the medial region might generate anisotropic cell shapes. On the other hand, the amplitude of the rotation and the contraction modes are similar for the medial and perimeter activity. However, the contraction response is slightly higher in the case of cell activity in the perimeter. These observations suggest that the net contraction and the degree of anisotropy of the final shape are significant observables for characterising a cell under this kind of cellular activity.

\section{Active cell embedded in a tissue}  \label{sec.tissue}

At a tissue level, due to the confluent property, competition between cells takes place. To study the response of a tissue when one of the cells in the tissue actively contracts, and to determine if the perimeter and medial activities produce different geometrical responses, we consider three numerical approaches. First, in the case of small contractions, we look for the steady solution of Eqs.\ \eqref{eq.variationaldyn}, which correspond to minimising the energy functional \eqref{eq.Evertex}. The linearised equations allow us to identify to main results, independently of the intensity of the contraction. Secondly, we solve the temporal evolution of the full non-linear equations Eqs.\ \eqref{eq.variationaldyn} for quite intense contractile activities, which gives besides the final state, the temporal scale needed to reach it. Finally, we consider tissues plasticity in the dynamics, meaning that overstretched cells  reconstruct their apical actomyosin network resulting in adaptation of the reference areas and perimeters $A_{0c}$ and $P_{0c}$. In all cases, we use periodic boundary conditions and to gain statistical accuracy, for each tissue, we run many simulations, each time applying the contraction to a different cell, which is monitored to measure the change in its geometry. We analyse two geometrical variables: the net contraction $R_c$ and the change on anisotropy $Q_c$, of the final state with respect to the relaxed state of the cell, defined as
\begin{align}
R_c &= \frac{A_c^{(\text{relax})} - A_c^{(\text{final})}}{A_c^{(\text{relax})}},\\
Q_c &= \frac{b_c^{(\text{relax})}}{a_c^{(\text{relax})}}-\frac{b_c^{(\text{final})}}{a_c^{(\text{final})}},
\end{align}
where $a_c$ and $b_c$ correspond to the principal semi-axes of the ellipse that better fits the cell $c$, with $a>b$, calculated as in Ref.\ \cite{perezinstability}.
Throughout this section, time is measured in the natural units $\hat{t}=1/(a^2K_A)$ and, whenever necessary, units are fixed such that $K_A=1$ and $a=1$. Finally, the time step used for the numerical integration of the dynamical equations is $dt=0.01$.

\subsection{Linear response} \label{sec.linear}
For this analysis, we consider tissues composed of 48  cells in a box of size $L_x = 6\sqrt{3}a$ and $Ly =12 a$ with periodic boundary conditions. Two types of tissues are considered. 
 i) A regular tissue made of perfectly hexagonal cells of side $a$, adding Gaussian noise of width $a/10$ to the vertex positions of the regular tessellation, which allows introducing disorder while keeping the hexagonal geometry. ii) An irregular tissue, built as Voronoi cells, where the positions of $N=48$ central points are generated by a Montecarlo simulation of hard disks in a box of equal size as for the regular tissue. The diameter of the disks governs the degree of dispersion of the cells and we consider an area fraction $\phi=0.71$, below the freezing transition, to obtain a disordered tessellation with moderate dispersion \cite{perezinstability}. This irregular tissue  presents a larger disorder, with  variability in the number of sides, as well as in the reference areas and perimeters. For both tissues, the generated lattice is considered as the initial configuration from where we calculate the initial values of the reference areas and perimeters, $A_{0c}$ and $P_{0c}$, used in the model.
Finally, given the initial positions,  we first solve Eqs.~\eqref{eq.variationaldyn} without considering activity term for various values of $K_P$ and $J$ for a relaxation time $t_{\text{relax}} = 1$, to reach equilibrium configurations.

\begin{figure}[ht!] 
 \includegraphics[width=.9\linewidth]{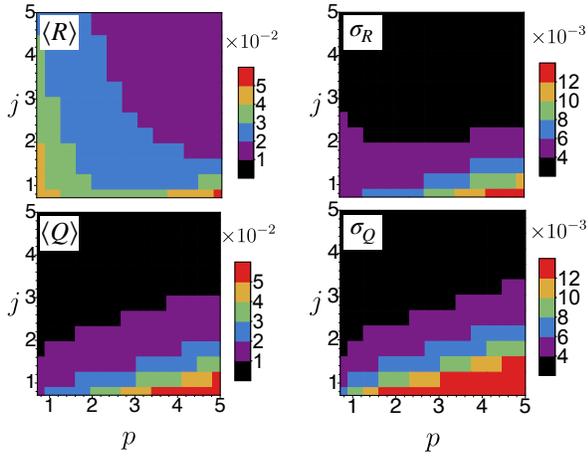} 
\caption{Analysis of the cellular linear response on a tissue composed of 48 irregular cells, where the averages and standard deviations are obtained by alternately applying a medial contraction with $\alpha_A=0.1$ and $\alpha_P=0$ over the 48 cells of the tissue. Mean values and standard deviations of the net contraction  $R$ (top) and change of anisotropy $Q$ (bottom) of the cells with respect to the relaxed configuration ($t_\text{relax}=1$), respectively.}
\label{fig.svdA0des}
\end{figure}

\begin{figure}[ht!] 
 \includegraphics[width=.9\linewidth]{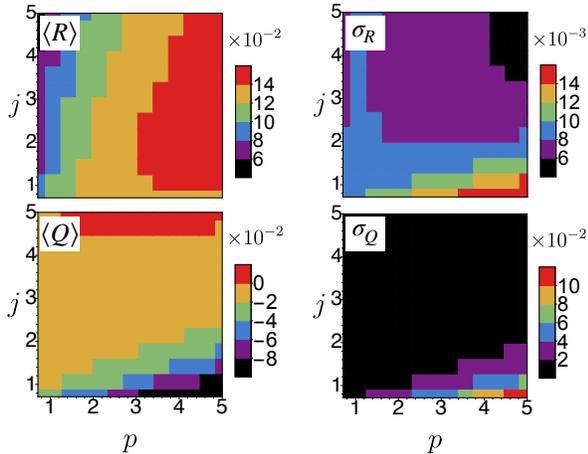} 
\caption{Same as in Fig.\ \ref{fig.svdA0des}, when a perimeter contraction with $\alpha_A=0$ and $\alpha_P=0.1$ is applied.}
\label{fig.svdP0des}
\end{figure}

On the relaxed tissue, we modify the value of $A_{0c}$ or $P_{0c}$ to a randomly selected cell and we write the set of linear equations $\derpar{E}{\VEC r_i}=0$ that determine the equilibrium positions, which are linearised considering small values of $\alpha_{A/P}$ and small displacements. The resulting matrix is ill-conditioned, needing the use of the Singular Value Decomposition (SVD) method. Figures \ref{fig.svdA0des} and \ref{fig.svdP0des} show the results for the irregular tissue, with $\alpha_A=0.1$ (and $\alpha_P=0$) and $\alpha_P=0.1$ (and $\alpha_A=0$), respectively. Statistical values are calculated by alternately applying the contraction to the whole set of 48 cells. We obtain a clear distinction in the linear response depending on the localisation of the activity. We find that the net contraction is bigger ($\sim$5 times) in the case of inner perimeter activity. Also, medial activity tends to elongate cells ($Q_c >0$), while inner perimeter activity tends to make the cells more isotropic ($Q_c <0$). The standard deviation of the net contraction, for both kinds of activity, is one order smaller than the mean value. Interestingly, on the contrary, for the change of anisotropy, it is of the same order. This reflects the high (low) dependence of the initial geometrical condition on the change of anisotropy (net contraction). The same results are obtained when considering the regular tissue with Gaussian noise on the positions.

\subsection{Non-linear dynamic response}

For larger deformations, we perform numerical simulations, solving the dynamical equations \eqref{eq.variationaldyn}, using ordered and disordered tissues of 494 cells arranged in a box of size $L_x = 19 \sqrt{3}a$ and $L_y=39a$ with periodic boundary conditions. Again, regular and irregular tissues are built as in Sect.\ \ref{sec.linear}. 

\begin{figure}[ht!] 
 \includegraphics[width=1\linewidth]{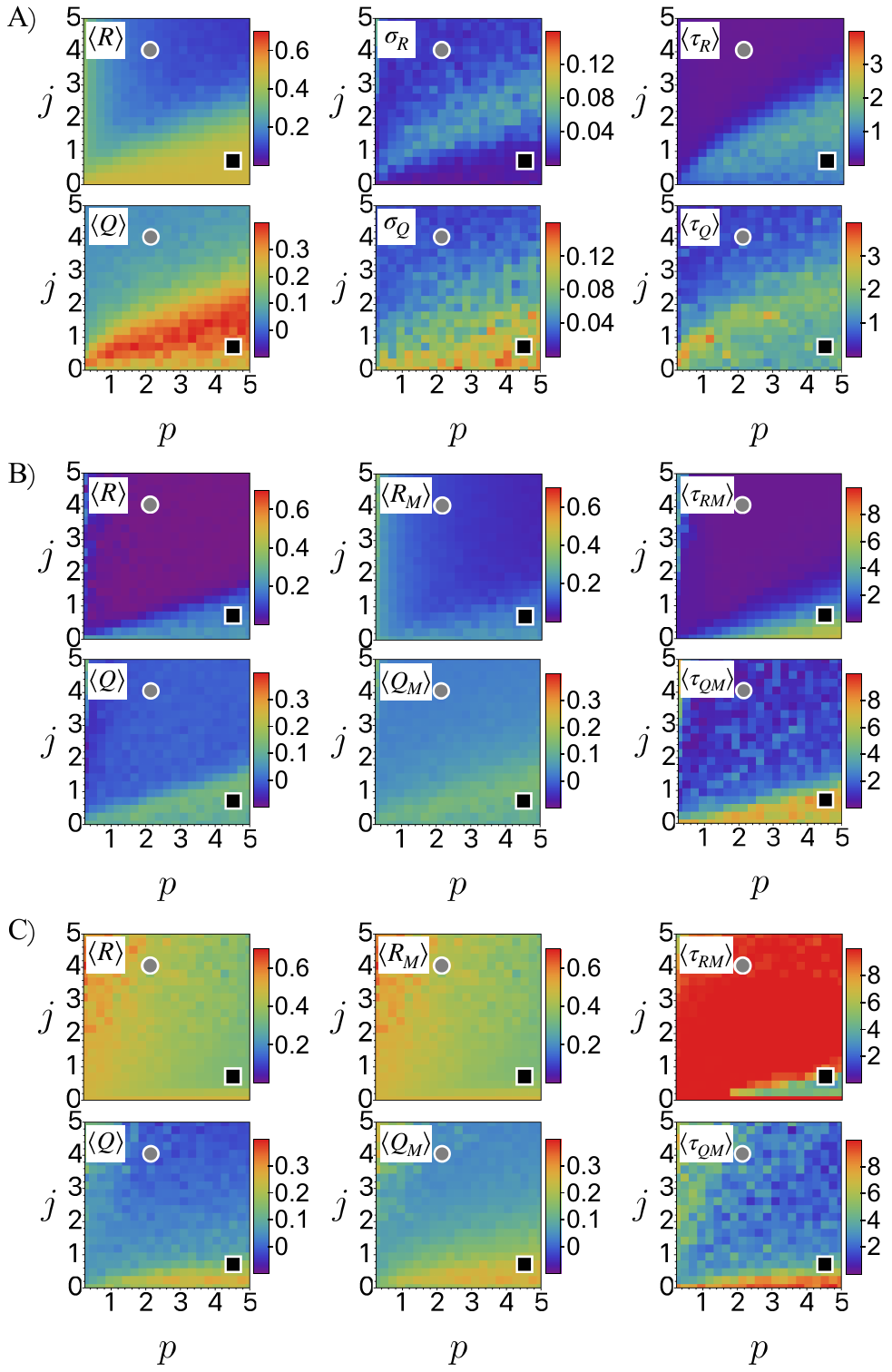} 
 \caption{Analysis of the cellular response to medial contraction with $\alpha_A=0.5$ and $\alpha_P=0$ obtained by  simulating  tissues of $494$ irregular cells up to $t_\text{final}=10$. A) Mean value (left), standard deviation (middle), and characteristic time (right) of the net contraction $R$ (up) and change of anisotropy $Q$ (bottom) of the contractile cells, respectively. B) and C)  Mean value (left), average of the maximum values (middle), and  time to reach the maximum values (right) of the net contraction $R$ (up) and change of anisotropy $Q$ (bottom) of the contractile cells, respectively. 
 For the statistical analysis, for each value of $p$ and $j$, the contraction was applied to 25 cells chosen at random. Times are measured in units of $\hat{t}= 1/(a^2K_a)$. The temporal evolution for selected values of parameters, shown with a grey circle and black square are shown in Fig.\ \ref{fig.A0c75}.
 In A) no plasticity is considered, while in B) $\nu_A=1$ and $\nu_P=0$, and in C) $\nu_A=0$ and $\nu_P=1$.}
\label{fig.disorderA0}
\end{figure}

\begin{figure}[ht!] 
\includegraphics[width=1\linewidth]{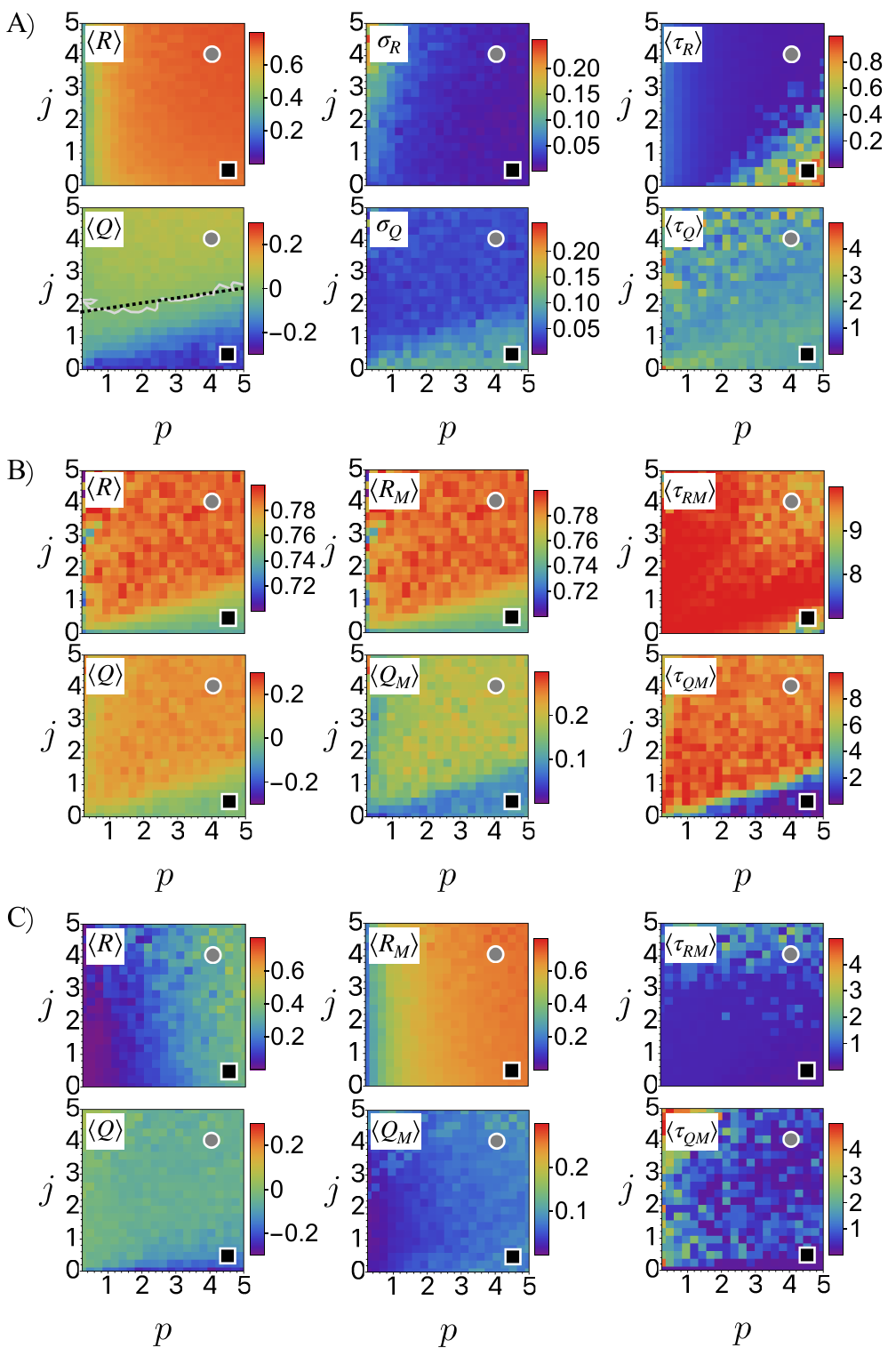} 
\caption{Same as in Fig.\ \ref{fig.disorderA0}, when perimeter contractions with $\alpha_A=0$ and $\alpha_P=0.5$ are applied. The white noisy curve in $\langle Q\rangle$ is the contour at which $\langle Q\rangle =0$; the black-dashed line corresponds to its linear approximation.
The temporal evolution for selected values of parameters, shown with a grey circle and black square are shown in Fig.\ \ref{fig.P0c75}.
In A) no plasticity is considered, while in B) $\nu_A=1$ and $\nu_P=0$, and in C) $\nu_A=0$ and $\nu_P=1$.}
\label{fig.disorderP0}
\end{figure}

We solve Eqs.~\eqref{eq.variationaldyn} for various values of $K_P$ and $J$. For all the simulations, we initially perform a relaxation phase, for $t_{\text{relax}} = 5$, to obtain the stationary configuration of the tissue. This configuration depends on the values of $j$ and $p$. Then, we study the system up to $t_\text{active}=10$.
Figures \ref{fig.disorderA0}-A) and \ref{fig.disorderP0}-A) correspond to the disordered tissues, showing the mean values and standard deviations  of $R$ and $Q$ at $t_{\text{final}}=10$, both with respect to the relaxed state  at $t_\text{relax}=5$. The statistical properties are obtained by sampling over  25 cells chosen at random, where the medial and perimeter contractions are applied, with $\alpha_A=0.5$ (and $\alpha_P=0$) and $\alpha_P=0.5$ (and $\alpha_A=0$), respectively.   Similar results are obtained for the regular tissue.

Again, it is obtained that perimeter activities generate final cellular shapes that are more isotropic and with greater levels of contractions as compared to the media activity, confirming that the results obtained with single cells and the linear dynamics. 
 By fitting an exponential function to the temporal evolution of $R_c$ and $Q_c$ for each of cells, we obtain characteristic relaxation times $\tau_R$ and $\tau_Q$, respectively, for each point in the $p$-$j$ parameter space. We find that, on average, for both activities, the relaxation of the change of anisotropy is slower than for the contraction. Also, the final net contraction is achieved faster with perimeter activity than with medial activity.

\subsection{Plasticity}
As mentioned, tissues present plastic behaviour, where for sustained applied stress, yielding takes place. At the macroscopic level, it is described with active gels models \cite{backouche2006active,joanny2009active}, where viscoelastic terms are included. In the vertex models, we include plasticity by allowing the reference areas and perimeters to evolve, according to the present value of the deformation. Namely, we add to the model the following dynamical equations
\begin{align}
\frac{d A_{0c}}{dt} &= -\nu_A\left( A_{0c}-A_{c}\right),
\label{eq.plastA0}\\
\frac{d P_{0c}}{dt} &= -\nu_P \left( P_{0c}-P_{c}\right),
\label{eq.plastP0}
\end{align}
where $\nu_A$ and $\nu_P$ give the plastic relaxation rates, which are related to the reconstruction time of the actomyosin network.
 
\begin{figure}[ht!] 
 \includegraphics[width=1\linewidth]{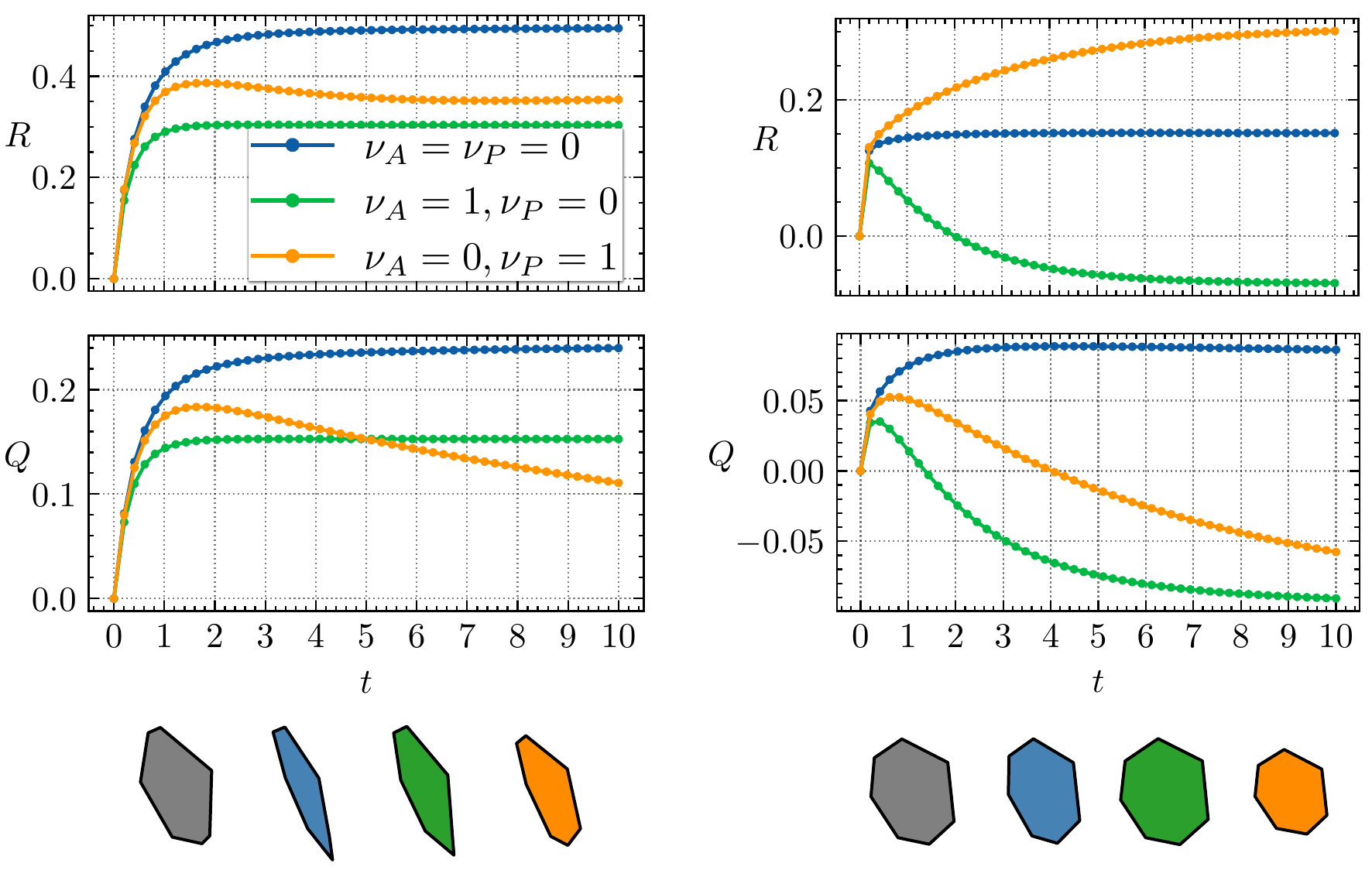} 
\caption{Temporal evolution of the geometrical variables for a selected cell in an irregular tissue, with parameters defined by the grey circle (left) and by the black square (right) of Fig.\ \ref{fig.disorderA0}. Net contraction $R$ (up) and change of anisotropy $Q$ (bottom), both measured with  respect to the relaxed configuration at $t_{\text{relax}}=5$ after a medial contraction with $\alpha_A=0.5$ and $\alpha_P=0$ is applied. Three cases are considered: no plasticities (blue), with plasticity in the medial region (green), and with plasticity in the perimeter  (orange). At the bottom, we show the cell in the initially relaxed configuration (grey) and the final configurations  at $t_\text{final}=10$ for each case, following the colours of the legend.}
\label{fig.A0c75}
\end{figure}

\begin{figure}[ht!] 
 \includegraphics[width=1\linewidth]{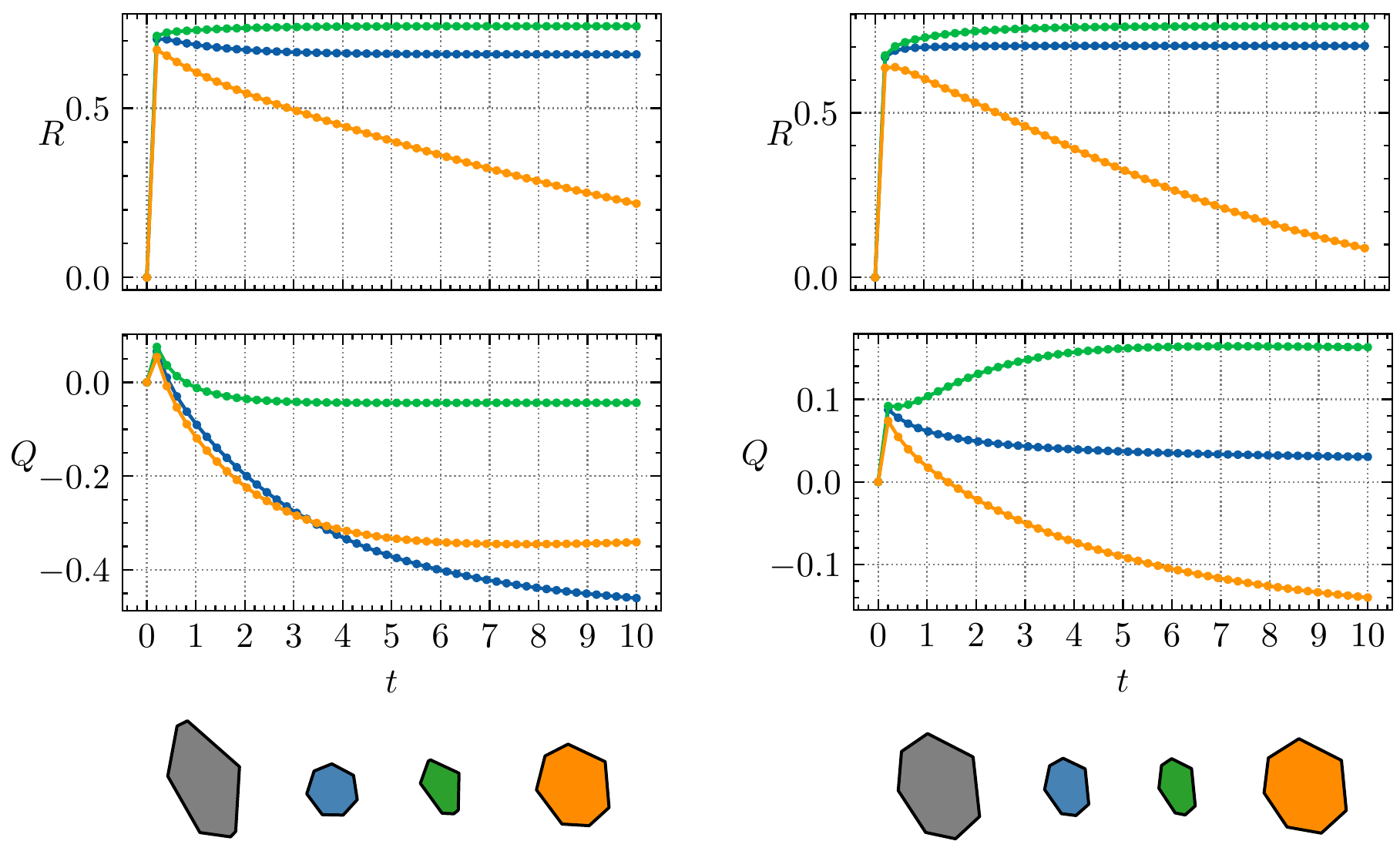} 
\caption{Same as in Fig.\ \ref{fig.A0c75} for the system parameters defined by the grey circle (left) and by the black square (right) of Fig.\ \ref{fig.disorderP0}, where a perimeter contraction with $\alpha_A=0$ and $\alpha_P=0.5$ is applied.}
\label{fig.P0c75}
\end{figure}

 Again, we simulate for $t_{{\text{final}}}=10$. Figures \ref{fig.A0c75} and \ref{fig.P0c75} show the comparison of the temporal evolution of $R_c$ and $Q_c$ for a given cell in the irregular tissue when the plastic evolution of the area and perimeter are switched on and off. There is a rich phenomenology, and when plasticity is switched on, the two geometrical quantities may present non-monotonic behaviours. For example, an important overshooting of the anisotropy is seen Fig.\ \ref{fig.A0c75}-right.
 
For a more systematic analysis, we perform a statistical analysis considering the temporal evolution for irregular tissues, where 25 cells are chosen at random to apply the contraction. Three quantities are computed: i) the mean value of the net contraction and change on anisotropy with respect to the relaxed configuration, both at $t_{{\text{final}}}=10$, $\langle R\rangle$  and $\langle Q\rangle$, respectively ; ii) the average maximum values of the two previous quantities, $\langle R_\text{max}\rangle$  and $\langle Q_\text{max}\rangle$; and iii) the average time at which the maximum values are achieved, $\langle \tau_{R\text{max}}\rangle$ and $\langle \tau_{Q\text{max}}\rangle$. The results are shown in Figs.\ \ref{fig.disorderA0}-B) ($\alpha_A=0.5, \nu_A=1$), \ref{fig.disorderA0}-C) ($\alpha_A=0.5, \nu_P=1$), \ref{fig.disorderP0}-B) ($\alpha_P=0.5, \nu_A=1$), and \ref{fig.disorderP0}-C) ($\alpha_P=0.5, \nu_P=1$).  

 We find that when the non-active equilibrium parameter is the plastic one, then the evolution of the cellular contraction is a monotonically increasing function, on average, and reaches higher values than in the non-plastic model. Instead, when the active equilibrium parameter is the plastic one, the evolution of the cellular contraction has a maximum at a short time ($t\sim 1$) and then decays, on average. Even the maximum values are smaller than in the non-plastic model.
For the change of cellular anisotropy, in the case of medial activity, we obtain smaller values than in the non-plastic model, for both kinds of plasticities. There are even some points in the $p$-$j$ space with negative values, on average. Instead, in the case of perimeter activity, this quantity increases, on average, for both kinds of plasticities. However, when the plasticity is in the equilibrium area, most of the $p$-$j$ space has positive values of anisotropy, while when the plasticity is in the equilibrium perimeter, most of the $p$-$j$  space has negative values of anisotropy. 
All this information tell us that significant contractions can also be achieved with medial activity if the process is slow enough and the perimeter region is plastic. Also, that the efficiency found earlier for the net contraction achieved with perimeter activity can increase in slow processes with plasticity in the medial region, at expense of elongating slightly the cells.

\section{Comparison with experiments} \label{sec.experiments}
The contraction modelling, in the medial and perimeter regions, was applied to describe the active contractile pulses that took place within the EVL epithelium during the developmental time that extends from 48 to \SI{59}{hpf} in \textit{A.\ nigripinnis}. From the experiment, it was noticed that the fertilised fish egg, composed of a blastoderm cup on top of a yolk cell, has an approximately spherical shape with radius $R=\SI{590}{\micro\meter}$. The blastoderm, sited at the animal pole, is composed of an epithelial tissue of thickness $\sim R/100$ composed of 68 cells, initially covering $\sim 12\%$ of the spherical surface (see Fig.\ \ref{fig.huevo_c22}-A)). The mean cell edge $l_0$ is around $\SI{50}{\micro\meter}$. See Ref.~\cite{reig2017extra} for additional details and the full experimental protocol. Throughout the time series of analysis, 16 epithelial cells, denoted in red colour in see Fig.\ \ref{fig.huevo_c22}-A), suffer active contraction pulses. An example of one of such events is shown in see Fig.\ \ref{fig.huevo_c22}-C) in which the mild contraction of an EVL cell is evident and the concomitant changes in the cell area, anisotropy and perimeter are quantified in see Fig.\ \ref{fig.huevo_c22}-B). Notably, after the contractile pulse, the cell recovers its area and perimeter, implying that the contraction is followed by an expansion phase. We noticed that other active epithelial cells showed similar behaviour reported previously. Concomitantly with contractile pulses, the cells rearrange decreasing their heights and, consequently, inducing an approximately linear expansion of the entire tissue. Accordingly, by the end of the time period, the covered surface has reached $\sim 15\%$. Experimental data include information regarding the three-dimensional vertex positions of every epithelial cell every $\SI{0.2}{\hour}$.  In numerical simulations, to consider the spherical geometry of the egg, we include the restriction $c=R/\vert\VEC r_i\vert-1$, using the Lagrange method, up to $O(dt)$ in the solution of Eqs.~\eqref{eq.variationaldyn}. For the activity of the pulses and the rearrangement of the tissue, we initially propose a model in which the evolution of the equilibrium areas and perimeters are given by

\begin{align}
\frac{d A_{0c}}{dt} &= -\nu_A\left( A_{0c}-A_{c}\right) + f_{A_c}(t) + A_{0c}(t=0)m_A,\\
\frac{d P_{0c}}{dt} &= -\nu_P\left( P_{0c}-P_{c}\right) + f_{P_c}(t)+P_{0c}(t=0)m_P,
\label{eq.plastAP01}
\end{align}
where $ f_{A_c}$ and $ f_{P_c}$ are the active functions that describe the pulses, and $m_A$ and $m_P$ account for the expansion due to the decrease of the cellular height. $A_{0c}(t=0)$ and $P_{0c}(t=0)$ correspond to the experimental values of the area and perimeter of each cell, at the beginning of the time series analysis. We consider $J=0$ for simplicity; therefore the initial state is considered as an equilibrium configuration and no relaxation is needed. The active functions are built as the concatenation of two semi-cycles of  sine functions, starting with a negative semi-cycle at $t_0$, with a duration of $\delta_1$, and following with a positive semi-cycle that lasts $\delta_2$. The amplitudes of the negative part ($C_1$), associated with rates of destruction of equilibrium material, are related to the amplitudes of the positive part ($C_2$), associated with rates of creation of equilibrium material, such that the total integral of the active functions is zero. 

Through the modelling, we work with a portion of the tissue composed of 39 cells excluding epithelial cells located at the border, since full tracking of them was not available for the entire time of analysis \cite{reig2017extra}. First, we simulate the pre-epiboly of the tissue with the evolution of the vertices of the active cells and those on the border of the analysed portion given by the experimental measures.  Hence, only the vertices of the non-active cells (denoted in green in Fig.\ \ref{fig.huevo_c22}-A)) that do not belong to active cells are free to move following Eqs.~\eqref{eq.variationaldyn}. Optimising a function that compares the cell areas and perimeters obtained in the numerical simulations with the experimentally measured values, we obtain an initial set of model parameters $\left[K_Al_0^2,K_P,\nu_A,\nu_P,m_A,m_P\right] = \left[\SI{5.75}{\hour^{-1}},\SI{5.16}{\hour^{-1}},\SI{0.72}{\hour^{-1}},\SI{1.14}{\hour^{-1}},\SI{0.05}{\hour^{-1}},\SI{0.01}{\hour^{-1}}\right]$.

\begin{figure}[ht!] 
 \includegraphics[width=1\linewidth]{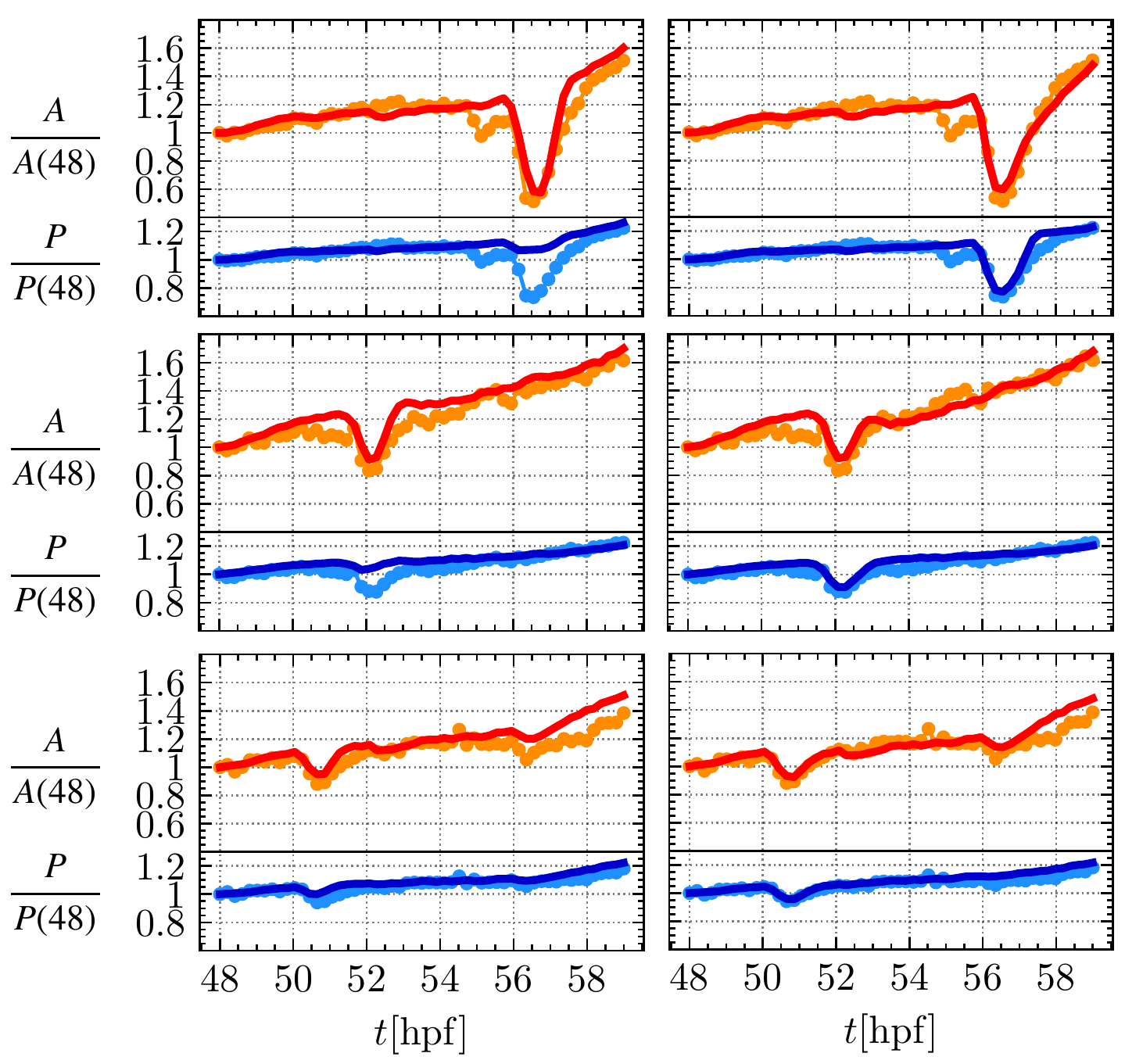} 
\caption{Geometrical evolution of three active cells with the parameters that result from optimisation processes. For each plot, in red (orange) is shown the numerical (experimental) area, and in blue (sky-blue) is shown the numerical (experimental) perimeter, both normalised to their initial values in the figures. Left: using medial activity. Right: using perimeter activity.
\label{fig.3act}}
\end{figure}

Then, we perform an active-cell-by-active-cell analysis for 3 cases, fixing $K_A$, $K_P$, $m_A$ and $m_P$ and letting the plasticity parameters $\nu_A$ and $\nu_P$ to change, since we expect the active events to be more sensible to their values.  For this analysis, we let free, as degrees of freedom, only the vertices that belong to the considered active cell, and the positions of all the other vertices are fixed to the experimental values. We search the best parameters that reproduce the active event when considering medial activity ($f_{A_c}$, but no $f_{P_c}$) or perimeter activity ($f_{P_c}$, but no $f_{A_c}$). In both cases, we fix initially, by looking at the experimental curves, the initial time of the active functions and the total duration of the pulse $\delta_T=\delta_1+\delta_2$.  We find that to avoid non-convex shapes of some cells and evolutions closer to the experimental ones, we might set $\nu_P=0$ and $m_P=0$, that is, there is neither plasticity nor rearrangement due to changes in the cell heights in the perimeter. 
Figure \ref{fig.3act} shows the results for the 3 active cells considered. We find that the perimeter activity matches the experimental results for both geometrical quantities, while medial activity is capable of diminishing the area as expected, but not the perimeter. This is directly related to more isotropic (anisotropic) shapes in the contraction achieved by using perimeter (medial) activity, as shown in the previous sections.

With the information gained after the initial simulations of the 3 cells, we performed parameter optimisations for all the active contractile events considering only perimeter activity. From those processes we obtain that $\delta_1\approx \delta_2$ on average. Hence,  $f_{P_c}$ is modelled as a complete cycle of a unique sine function, allowing to reduce the number of parameters in the optimisation.
With these considerations, the evolution of the equilibrium areas and perimeters are now given by
\begin{align}
\frac{d A_{0c}}{dt} &= -\nu_A\left( A_{0c}-A_{c}\right) + A_{0c}(t=0)m_A,\\
\frac{d P_{0c}}{dt} &= f_{P_c}(t),
\label{eq.plastAP02}
\end{align}
Finally, we perform a last optimisation considering the above simplifications obtaining curves similar to those of Fig.\ \ref{fig.3act}-right for all pulses. The global optimised parameters are $K_A l_0^2=\SI{6.25}{\hour^{-1}}$, $K_P=\SI{4.82}{\hour^{-1}}$, $\nu_A=\SI{3.20}{\hour^{-1}}$, and $m_A=\SI{0.05}{\hour^{-1}}$, where we fix $\nu_P$ and $m_P$ to zero, and the  optimised parameters for each pulse are given in Table \ref{table.fit}.

\begin{table}[htb]
\begin{tabular}{c|c|c|c}
Cell number & $t_0 [\si{\hour}] $ & $\delta_T [\si{\hour}] $ & $C_1[\si{\micro\meter/\hour}]  $\\
\hline
1 & 0.40 & 3.00 & 0.19\\
2 & 3.80 & 7.00 & 0.10\\
3 & 2.00 & 2.00 & 0.44\\
4 & 0.20 & 1.60 & 0.38\\
5 & 3.60 & 2.10 & 0.53\\
6 & 3.80 & 2.00 & 0.45\\
7 & 0.00 & 2.00 & 0.33\\
8 & 3.60 & 3.60 & 0.27\\
9 & 3.50 & 1.80 & 0.62\\
10 & 4.00 & 4.00 & 0.12\\
11 & 2.00 & 1.20 & 0.43\\
12 & 6.30 & 2.20 & 0.56\\
13 & 3.20 & 1.60 & 0.75\\
14 & 6.00 & 2.00 & 0.32\\
15 & 6.80 & 2.50 & 0.67\\
\end{tabular}
\caption{Parameters characterising each pulse, where the perimeter active functions are given by $f_{P_c}= -C_1 \sin(2\pi(t-t_0)/\delta_T)$ for $0\leq t-t_0\leq \delta_T$. The start times $t_0$ and pulse durations $\delta_T$ are fixed by the experiments, while the pulse amplitudes $C_1$ are fitted to best reproduce the experimental evolution of the areas and perimeters.}
\label{table.fit}
\end{table}

\section{Discussion} \label{sec.discussion}
Our successive analysis (one isolated active cell,  an active cell embedded in a tissue with linear response and full non-linear dynamics) showed that apical constrictions described by the two-dimensional vertex model where the activity enter as modifications of the equilibrium parameters, present different geometrical characterisation that  depends on the cellular region taken as the active zone. When the inner perimeter of the cell is active, as in the purse-string model, resembling actomyosin ring-induced active stresses, cells contract more efficiently than when the medial region of the cell is active, as in the meshwork model. This feature is robust and maintained when plasticities are considered in the system. For the anisotropy, in the non-plastic case, when the inner perimeter of the cell is active, cells achieve circular shapes with the contraction, while when the medial region of the cell is active they reached elongated shapes. This feature, in contrast, does not hold when the medial region is plastic, in which case an active perimeter can generate elongated shapes.

The application of our modelling to the  active contractile pulses extending through blastula stages of the annual killifish \textit{A.\ nigripinnis} is highly illustrative. We were able to quantitatively fit the pulses of 15 cells with a reduced number of parameters (4 global parameters -- $K_A, K_P, \nu_A, m_A$ -- and 3 specifics parameters per cell -- $t_0$, $\delta_T$, $C_1$). Also, the analysis showed that the pulses, in which cells contract keeping roughly isotropic shaped, and thus reducing both their areas and perimeters, are better described with activity only in the perimeter of the cells and plasticity on the areas. We recall that by comparing the cell areas and perimeters, which are simple geometrical observables that are easily accessible, it is possible to discriminate between the two possible sources of activity and, also, to fit all the relevant parameters. 
The presented methodology provides a practical and quantitative perspective of how apical constrictions occur in other systems. Also, offers a method to measure different parameter used in the vertex or similar models in many biological systems. It remains an interesting perspective to relate the measured parameters to the relevant biophysical processes taking place in the cells and tissues.

\acknowledgments
This research was supported by the ANID (Chile) projects: Millennium Science Initiative Program-NCN19\_170 to R.S., M.C.L. and F.P.; FONDECYT 1180791 to R.S.; FONDECYT 1190806 to M.L.C., M.C. and R.S.; FONDECYT 11170761 to G.R.; PIA/ACT192015 and Millennium Institute ICN09\_015 to M.L.C and M.C.; FONDEQUIP EQM130051 and FONDAP 15150012 to M.L.C.; Climat-AmSUD CLI2020004 to G.R.


%

\end{document}